\documentclass[useAMS,usenatbib]{mn2e}

\usepackage{graphicx}

\newcommand{\Msun}{$M_{\odot}$}

\newcommand{\kms}{km\,s$^{-1}$}
\newcommand{\vs}{$v \sin i$}
\newcommand{\teff}{$T_{\rm eff}$}
\newcommand{\lgg}{$\log\,{g}$}

\title[The close binaries HD 22128 and HD 56495]
{Do the close binaries HD 22128 and HD 56495 contain Ap or Am stars?\thanks{Based on observations obtained at the Bernard Lyot Telescope (TBL, Pic du Midi, France) of the Midi-Pyr\'en\'ees Observatory, which is operated by the Institut National des Sciences de l'Univers of the Centre National de la Recherche Scientifique of France. }
}

\author[Folsom et al.]{C.P. Folsom$^{1}$\thanks{E-mail: cpf@arm.ac.uk}, G.A. Wade$^{2}$, 
N.M. Johnson$^{2}$\\
$^{1}$Armagh Observatory, College Hill, Armagh Northern Ireland BT61 9DG\\
$^{2}$Department of Physics, Royal Military College of Canada, P.O. Box 17000, Station `Forces', Kingston, Ontario, Canada, K7K 7B4}

\begin{document}

\date{Received: 2013; Accepted: 2013}

\pagerange{\pageref{firstpage}--\pageref{lastpage}} \pubyear{2013}

\maketitle

\label{firstpage}

\begin{abstract}
HD~22128 and HD~56495 are both double-lined spectroscopic binary systems with 
short orbital periods, which have been proposed to host magnetic Ap stars. 
Ap stars in short period binary systems are very rare, 
and may provide insight into the origin of magnetism in A-type stars.  
We study these two systems using high-resolution MuSiCoS spectropolarimetric 
data, in order to asses the presence of magnetic fields and study the 
atmospheric chemistry of the components.  
This represents the first modern magnetic measurements 
and careful spectroscopic analyses of these stars.  
We find no evidence of a magnetic field in any of the stars, 
with precise uncertainties on the longitudinal magnetic field of 50 and 80 G 
in the components of HD~22128, and 80 and 100 G in the components of HD~56495.  
We performed detailed abundance analyses of both stars in both systems, 
finding clear evidence of Am chemical peculiarities in both components of 
HD~22128, and in the brighter component of HD~56495, with overabundant iron peak 
elements and underabundant Sc and Ca.  
The less luminous component of HD~56495 is chemically normal.  
The atmospheric chemistry is consistent with the absence of magnetic fields, 
and consistent with the theory of Am star formation proposing that tidal 
interactions slow the rotation rate of the star, allowing atomic 
diffusion to proceed efficiently.  
\end{abstract}


\section{Introduction}

Among A and B-type stars, strong organised magnetic fields are rare, 
occurring in 5 to 10\% of these stars.  The chemically peculiar 
Ap and Bp stars appear to always have such magnetic fields \citep{Auriere2007}, 
while chemically normal stars, and other classes of chemically peculiar stars, 
appear to never possess such fields \citep[e.g.][]{Shorlin2002,Wade2006-alphaAnd,Makaganiuk2011-HgMn-magnetic-survey}.
An important unresolved question is why strong organised magnetic fields 
appear in a small subset of A and B stars, and are absent in the majority 
of A and B stars.  

An interesting, and possibly related, observation is that Ap stars are particularly rare in close binary systems.  
Modern surveys \citep{Gerbaldi1985-Ap-binary-frequency,Carrier2002-binarity_in_Ap} 
find Ap stars exhibit about the same binary frequency as normal A-type stars 
\citep[$47\pm5$\%,][]{Jaschek1970-old-binary-frequency}.  
However they also identify a clear lack of Ap stars in short-period 
(i.e. $P_{\rm orb} \leq 3.0$ days) systems \citep{Carrier2002-binarity_in_Ap}. 
This almost certainly provides a clue about the origin of 
magnetic fields in A and B stars.  Consequently, Ap/Bp stars in close binary 
systems are particularly important systems for detailed investigations.
The Binarity and Magnetic Interactions in Stars (BinaMIcS) collaboration 
has recently been founded, with the investigation of 
such systems as one of the primary goals.  

While Ap stars appear relatively rarely in short period binaries, 
other types of chemically peculiar stars occur in such  
systems relatively frequently. 
Am stars occur more commonly in binary systems than normal A stars, 
and this trend continues to short period binaries 
\citep{Abt1961-binary-Am,Abt1985-binary-Am-2,Carquillat2007-survey-Am-SB}. 
Thus binary interactions do not inhibit the formation of chemical peculiarities.  
Indeed, a popular hypothesis is that tidal interactions in binary systems slow 
the rotation rates of Am stars, thereby reducing mixing from meridional 
circulation, and allowing atomic diffusion to proceed efficiently 
\citep{Michaud1983-Am-diffusion-massloss-rotation,Abt1985-binary-Am-2}. 
Atomic diffusion is commonly thought to be the physical process giving 
rise to chemical peculiarities in both Ap and Am stars 
\citep{Michaud1970-diffusion,Michaud1981-diffusion_magneticApBp}.  

Systems containing an Ap star and another A star are particularly 
interesting, since they provide the opportunity to study the incidence 
of Ap stars in close binaries, and also the evolution of magnetic fields and 
chemical peculiarities in two similar stars.  
A careful literature review reveals there are only five known 
or proposed cases of an Ap star in a SB2 binary with a 
main sequence A star.
HD~55719 was the first well established system, studied by 
\citet{Bonsack1976-hd55719-Ap-binary}, although as a southern object and 
has not received much modern attention.  
HD~98088 was established as a binary by \citet{Abt1953-hd98088-vr} and 
\citet{Abt1968-hd98088-vr-orbit}, and as system with a magnetic Ap star 
by \citet{Babcock1958-magnetic-catalog}.  HD~98088 was recently studied 
in detail by \citet{Folsom2013-HD98088-Ap-binary} who confirm the primary 
is an Ap star, find the secondary is an Am star, and present a dipole model 
of the magnetic field of the primary as well as a detailed abundance analysis.  
The stars HD~5550, HD~22128, and HD~56495 all show SB2 spectra with 
an A-type primary \citep[e.g.][]{Carrier2002-binarity_in_Ap}, however 
the magnetic nature of the proposed Ap component has yet to be 
definitively established.  
Two other SB2 systems containing an Ap star are known: 
HD~135728 \citep{Freyhammer2008-new-Ap+1binary} and HD~59435 
\citep{Wade1999-hd59435-Ap-binary}, however these systems have 
giants of spectral type G8 as their primary components. 
All of these stars represent potential targets for intensive   
observation by the BinaMIcS project.  
This paper focuses on HD~22128 and HD~56495, both of which have 
been poorly studied but are potentially very interesting systems.

HD 22128 was first identified as possibly being chemically peculiar by 
\citet{Olsen1979-photometric-classification} based on Str\"omgren photometry.  
\citet{Abt1979-classification-cp} classified the star as an Ap star 
(A9 IVp Sr,Eu,Mn st., Ca wk) based on classification resolution spectra. 
\citet{Renson1991-Ap-Am-catalogue} and \citet{Renson2009-catalogue-updated} 
included HD 22128 in their catalogue 
of chemically peculiar stars as an Ap star (A7 SrEuMn), but also include a note 
that the star may be an Am star.  There are no magnetic measurements in 
the literature for HD 22128.   \citet{Carrier2002-binarity_in_Ap} 
identified the star as an SB2, and they derived precise orbital parameters 
which we reproduce here in Table \ref{orbital-param}.  
\citet{Sowell2001-binary-speckle} made speckle observations of the system, 
but found no second component down to separations of $\sim$0.2'', as long as 
the difference in magnitude between the components is less than 2 mag. 
\citet{Horch2011-binary-speckle} also made speckle observations 
of the system, finding a similar 0.2'' limit for a difference in 
magnitude of less than 4.5.  This is consistent with the separation found by 
\citet{Carrier2002-binarity_in_Ap}, both for their estimated orbital inclination, 
and for all but the most extreme inclinations.

HD 56495 was first reported to be a magnetic A3p star by 
\citet{Babcock1958-magnetic-catalog}, who obtained a marginal magnetic 
detection of $B_{l} = +570 \pm 200$ G, and a non-detection of $+210 \pm 220$ G.
No magnetic measurements have been reported in the literature since. 
\citet{Bertaud1959-Ap-Catalogue} included it as an A3p Sr star in their 
catalogue of A stars with spectroscopic peculiarities.  
However HD~56495 was classified as an Am star by 
\citet{Bertaud1967-Am-classifications} (with spectral types A2 from Ca K 
and F2 from metallic lines), based on spectroscopy.  
\citet{Renson1991-Ap-Am-catalogue} and \citet{Renson2009-catalogue-updated} 
included HD~56495 in their catalogue 
as an Am star, with a note that a magnetic field had been measured by 
\citet{Babcock1958-magnetic-catalog}.  
\citet{Bychkov2003-magnetic-cat,Bychkov2009-magnetic-cat-update} 
also include HD~56495 in their catalogue of stellar magnetic fields 
as an Am star, with an effective magnetic field of $430 \pm 238$ G based 
on the measurement from \citet{Babcock1958-magnetic-catalog}.  
HD~56495 was suggested to be SB2 by \citet{Hynek1938-survey-SB2}, 
though the first orbital parameters were found by 
\citet{Carrier2002-binarity_in_Ap}, which we also reproduce 
in Table \ref{orbital-param}.

\begin{table}
\centering
\caption{Best fit orbital parameters from \citet{Carrier2002-binarity_in_Ap}. }
\scriptsize
\label{orbital-param}
\begin{tabular}{lcc}
\hline \hline
                      & HD 22128               & HD 56495 \\
\hline
$P$ (d)               & $5.085564 \pm 0.000070$& $27.37995 \pm 0.00080$ \\
$T_0$ (HJD -2,400,000) & $50116.7656 \pm 0.0043$& $48978.40 \pm 0.23$ \\ 
$e$                   & $0.00$ (fixed)         & $0.1651 \pm 0.0097$ \\
$V_0$ (\kms)          & $15.30 \pm 0.21$       & $-7.57 \pm 0.35$ \\
$\omega$ ($\degr$)    &  -                     & $224.7 \pm 3.2$ \\
$K_{\rm A}$ (\kms)     & $68.40 \pm 0.37$       & $44.30 \pm 0.74$ \\ 
$K_{\rm B}$ (\kms)     & $73.69 \pm 0.55$       & $57.75 \pm 0.81$ \\ 
$M_{\rm A} \sin^3 i$ (\Msun)& $0.786 \pm 0.012$ & $1.641 \pm 0.055$ \\
$M_{\rm B} \sin^3 i$ (\Msun)& $0.729 \pm 0.010$ & $1.259 \pm 0.044$ \\
$a_{\rm A} \sin i$ ($10^6$ km)& $4.784 \pm 0.026$&$16.45 \pm 0.27$ \\
$a_{\rm B} \sin i$ ($10^6$ km)& $5.153 \pm 0.038$&$21.44 \pm 0.30$ \\
\hline \hline
\end{tabular}
\end{table}

\section{Observations}
\label{observations}

Observations of HD 22128 and HD 56495 were obtained with the 
Multi-Site Continuous Spectroscopy (MuSiCoS) spectropolarimeter, 
attached to the the T\'elescope Bernard Lyot at the Observatoire du Pic du Midi, France.
MuSiCoS, which has since been decommissioned, consists of a Cassegrain 
mounted polarimeter unit \citep{Donati1999-musicos-pol}, attached by optical 
fibres to a bench mounted cross-dispersed \'echelle spectrograph 
\citep{Baudrand1992-musicos-spectrograph}.  
The instrument has a resolution of $R=35\,000$ with a wavelength range from 
4500 to 6500 \AA, and provides polarised Stokes $V$, $Q$ and $U$ spectra 
as well as total intensity Stokes $I$ spectra.  
Data reduction was performed with the ESpRIT \citep{Donati1997-major} reduction 
tool, which performs optimal 1D spectrum extraction, as well as the relevant 
calibrations.  The reduced spectra were continuum normalised by fitting a low 
order polynomial through carefully selected continuum points, then dividing the 
observation by the continuum polynomial.  

Four observations of HD 22128 were obtained in Stokes $V$ (providing Stokes 
$I$ as well) in Feb.\ 2004.  For HD 56495, four Stokes $V$ observations were 
also obtained in 2004, however one of the observations (9 Feb, 2004) 
had very low S/N, due to poor observing conditions,  
and is discarded in the subsequent analysis.  
The observations used are reported in Table \ref{observations_table}.

\begin{table*}
\centering
\caption[Summary of MuSiCoS observations]
{Summary of MuSiCoS observations. The UT date, the Heliocentric Julian Date,
and the peak signal to noise ratio are given.  For all observations of both stars, 
four subexposures of $800$s duration were used.  
The orbital phase was calculated from the ephemerides of \citet{Carrier2002-binarity_in_Ap}.  
The radial velocity and longitudinal magnetic field ($B_\ell$) 
were measured from LSD profiles.  }  
\begin{tabular}{rrrrrrrr} 
\hline\hline
\multicolumn{1}{c}{Date}& \multicolumn{1}{c}{HJD}& Peak &\multicolumn{2}{c}{Velocity (\kms)}& Orbital & \multicolumn{2}{c}{$B_\ell$ (G)}\\
             &(2,450,000+)&S/N &  Primary       & Secondary        & Phase   &  Primary     & Secondary       \\ 
\hline
\multicolumn{8}{c}{HD 22128} \\
\hline
 9 Feb, 2004 & 3045.408 & 110 & $+66.7 \pm 0.1$ & $-36.0 \pm 0.3$ & 0.8737 &$ -24 \pm 41$&$ +63 \pm 82$\\
10 Feb, 2004 & 3046.368 & 100 & $+79.2 \pm 0.1$ & $-50.4 \pm 0.3$ & 0.0623 &$ +70 \pm 48$&$ -28 \pm 84$\\
12 Feb, 2004 & 3048.333 & 100 & $-49.0 \pm 0.1$ & $+87.4 \pm 0.3$ & 0.4489 &$ +10 \pm 54$&$-167 \pm115$\\
14 Feb, 2004 & 3050.327 & 120 & $+55.5 \pm 0.1$ & $-25.1 \pm 0.2$ & 0.8408 &$ -45 \pm 27$&$ -59 \pm 57$\\
\hline
\multicolumn{8}{c}{HD 56495} \\
\hline
16 Jan, 2004 & 3020.567 &  81 & $ -3.6 \pm 0.5$ & $-12.3 \pm 0.6$ & 0.6324 &$ +28 \pm 88$&    blended  \\
 1 Feb, 2004 & 3037.475 &  98 & $+28.2 \pm 0.2$ & $-46.2 \pm 0.2$ & 0.2499 &$ +56 \pm 69$&$  -6 \pm 77$\\
 7 Feb, 2004 & 3042.572 & 100 & $+28.2 \pm 0.2$ & $-50.4 \pm 0.2$ & 0.4361 &$-137 \pm 80$&$-198 \pm119$\\
\hline\hline
\end{tabular}
\label{observations_table}
\end{table*}

\section{Magnetic measurements}
\label{magnetic-fields}

In order to investigate the presence of magnetic fields in HD~22128 and HD~56495, 
Least-Squares Deconvolution \citep[LSD,][]{Donati1997-major,Kochukhov2010-LSD} 
was applied to each of the calibrated and normalised spectra. 
LSD is a cross-correlation technique that produces an `average' line profile, 
with a much higher S/N than an individual line in the observation.  
The implementation of LSD by Donati was used, though the results was checked 
against our own implementation of the procedure, and found to be fully consistent.
Line masks were constructed using the atmospheric parameters and chemical 
abundances derived in Sect.\ \ref{Chemical Abundances}. 
Atomic data for the line masks, and predicted line depths, 
were extracted from the Vienna Atomic Line Database (VALD) 
\citep{Kupka1999-VALD}, using an `extract stellar' request.  

In HD~22128 both components have very similar parameters, and 
thus we can safely use the same mask for both.  In HD~56495, the components' 
temperatures and photospheric abundances differ somewhat.  
We therefore constructed separate line masks corresponding to the 
parameters of the two components of HD~56495.  
For HD~56495, we tested the influence of different line masks, including  
using the line mask of the primary for the analysis of the secondary, and 
varying the abundances used for the masks by our uncertainties in abundance.  
This resulted in an insignificant difference in the results, so ultimately 
we used a mask corresponding to the primary for the analysis of the primary, and 
a mask corresponding to the secondary for the analysis of the secondary. 

\begin{figure*}
\centering
\includegraphics[width=3.4in]{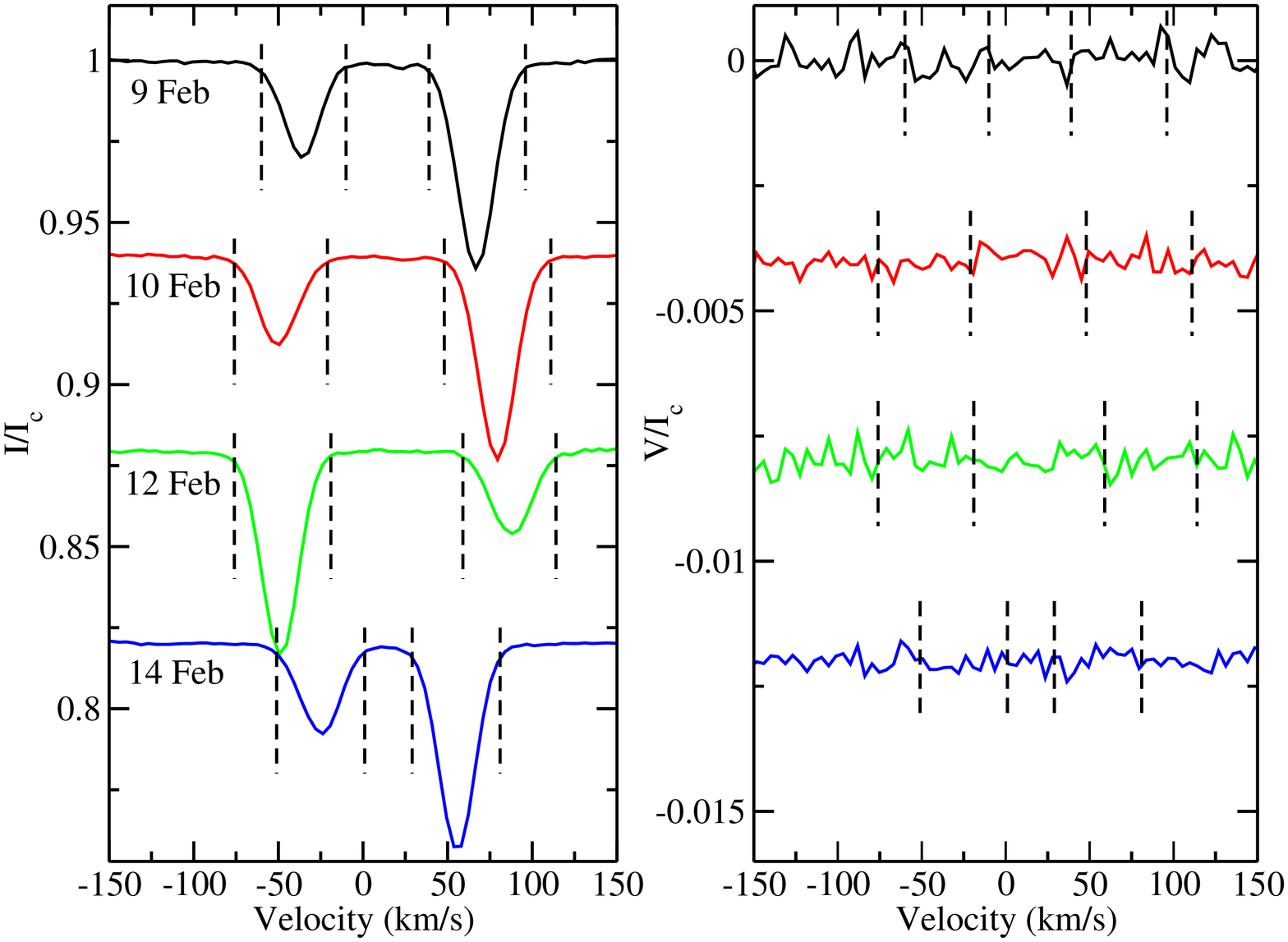}
\includegraphics[width=3.4in]{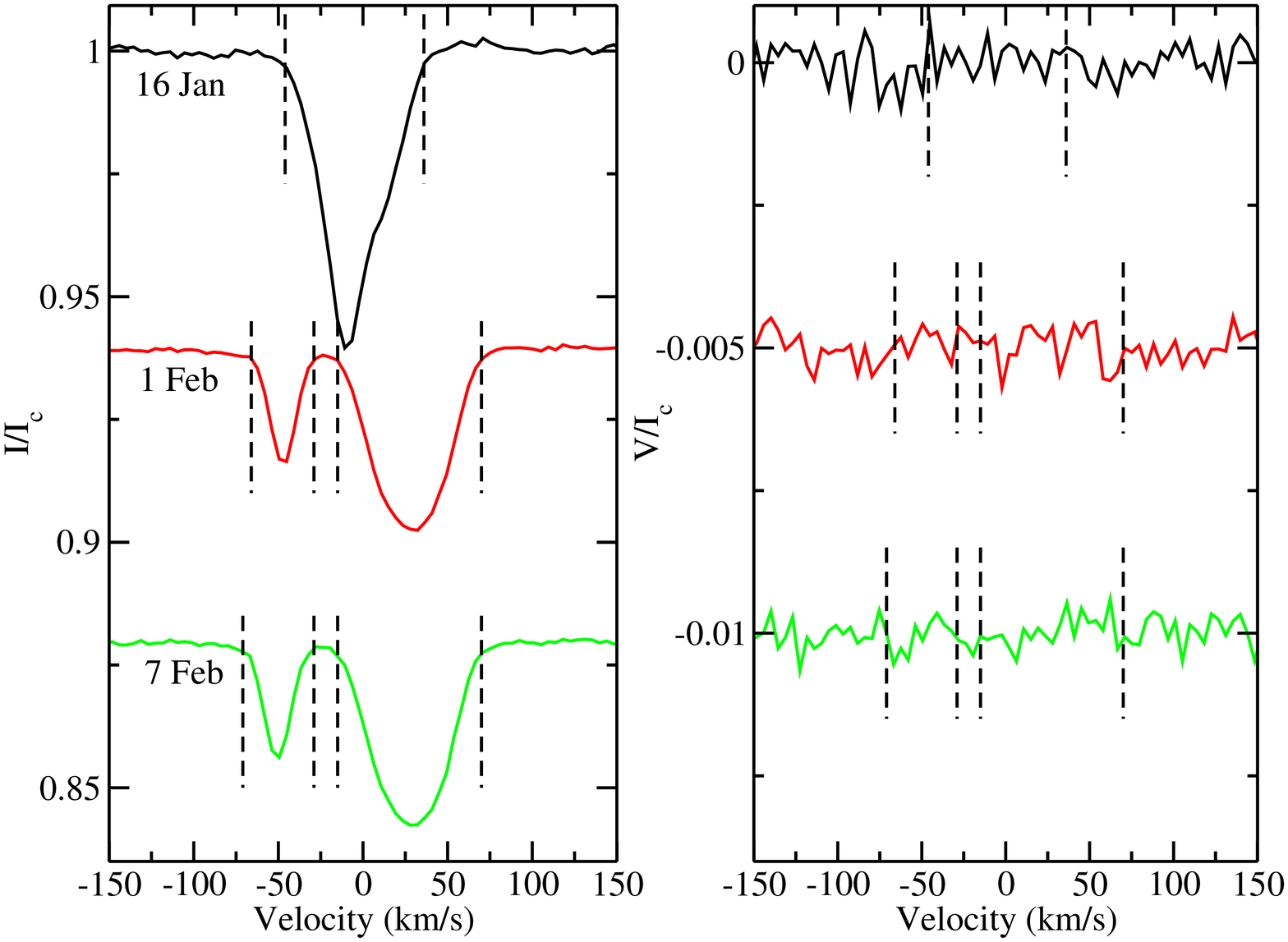}
\caption{LSD Stokes $I$ and $V$ profiles extracted from the observations of 
HD~22128 (left two frames) and HD~56495 (right two frames, for the LSD mask of the primary).  
The weaker lines from the secondary components of both systems are visible, 
however in one observation of HD~56495 the lines of the two components are 
completely blended.  
No polarisation signatures can be seen in $V$ in either in any observation 
of either system.  The LSD profiles are labelled by date, 
and are shifted vertically for clarity.  Vertical dashed lines indicated 
the integration used to measure longitudinal magnetic fields.  }  
\label{lsdprofs}
\end{figure*}

The amplitudes of the $I$ LSD profiles are normalised by the mean line depth, 
and the amplitudes of the $V$ LSD profiles are normalised by 
the product of the mean depth, wavelength, 
and Land\'e factor \citep{Kochukhov2010-LSD}.  
For HD~22128, the mean Land\'e factor is 1.159, the mean wavelength is 
542.0 nm, and the mean line depth is 0.323 of the continuum. 
For HD~56495 the mean Land\'e factors are 1.192 and 1.167, the mean 
wavelengths are 542.9 and 537.8 nm, and the mean line depths are 0.382 
and 0.324 of the continuum, for the primary and secondary masks respectively. 
The resulting LSD profiles are presented in Table \ref{lsdprofs}, 
for both Stokes $I$ and $V$. 

Radial velocities for the components of the two binary systems were measured 
from the LSD profiles, for each of our observations.  This was done by fitting 
a Gaussian line profile to the LSD line profile of the relevant component, 
through $\chi^2$ minimisation, and taking the centroid of the best fit Gaussian 
as the radial velocity. 
In order to estimate an uncertainty on the radial velocity this process was 
repeated for several different line masks, and different spacings between points 
in the extracted LSD profile (varying by $\sim 20\%$), and the full scatter 
between these runs was taken as the uncertainty.  
This attempts to sample the noise in the observation somewhat differently, 
and also account for any systematics introduced by choice of the line mask.  
The choice of line mask was the largest contributor to the uncertainty 
in radial velocity.  The only clear systematic trend observed was that using 
a solar abundance mask for the Am stars produced radial velocities 0.1 
or 0.2 km/s below the velocities from an Am mask.  It is possible an opposite 
trend occurs for the chemically normal HD~56495~B, but this is not clear due 
to slightly larger random errors.  The details of the Am abundances used in the 
mask and the line depth cutoff used for the mask had no clear systematic impact.  
The radial velocities are reported in Table \ref{observations_table}.

The Stokes $V$ LSD profiles were searched for a signal of a magnetic field, 
which would be produced by the longitudinal Zeeman effect.  We used the 
detection criterion of \citet{Donati1997-major}, in which the $V/I_{c}$ profile 
is searched for significant deviations from the null profile.  
A null profile is fit to the observed LSD profile, and the resulting $\chi^2$ 
is used to estimate the false alarm probability (FAP) of a deviation from the null.
A conservative FAP of $< 10^{-5}$ is used as the definite detection criteria of 
a magnetic field, and a marginal detection requires a FAP of $< 10^{-3}$. 
FAPs were measured for all LSD profiles and all of the values were below the 
marginal detection threshold.  

Despite the absence of magnetic detections in the $V$ LSD profiles, we can still 
measure longitudinal magnetic fields from the profiles and use these to place upper limits 
on the longitudinal field.  We measured the magnetic field using the first 
order moment method, described by \citet{Rees1979-magnetic-cog}. 
%\citet{Donati1997-major} and \citet{Wade2000-highPrecision-correctBz}.  
This involves integrating the (continuum normalised) LSD profiles $I/I_{\rm c}$ and 
$V/I_{\rm c}$ profiles about the centres-of-gravity ($v_{\rm 0}$) in velocity $v$: 
\begin{equation}
B_{\ell}=-2.14\times 10^{11}\ \frac{{\displaystyle \int (v-v_{\rm 0}) V(v)\ dv}}{\displaystyle {\lambda z c\ \int [1-I(v)]\ dv}},
\label{bz-equation}
\end{equation}
where the wavelength $\lambda$ (expressed in nm) and the Land\'e factor $z$ correspond 
to the weighting values used in computing the LSD profiles, and the measured longitudinal 
field $B_{\ell}$ is in gauss.  

Integration ranges for Eq.~\ref{bz-equation} were deduced by eye, to include the 
complete range of the line profile in $I$.  
Changing this integration range by 10\% typically changes the measured magnetic 
fields by $0.5\sigma$, and in some cases up to $1\sigma$, but in no cases does 
this convert a non-detection into a detection.  
The measured longitudinal magnetic fields are unaffected by the continuum 
of the companion in the SB2 spectra.  This is because the continuum that would appear 
in the numerator and denominator of Eq.~\ref{bz-equation} cancels out.  
However the uncertainty, based on propagating photon noise uncertainties through 
the equation, is larger for the SB2 spectrum.  

The measured longitudinal fields are presented in Table \ref{observations_table}, 
all of which are non-detections.  In one observation of HD~56495 the two components 
were completely blended, and a single effective longitudinal field was 
calculated for the full blended line.  
For HD~22128~A we find typical $1\sigma$ uncertainties of 50 G, and for HD~22128~B 
the typical uncertainties are 80 G.  For HD~56495~A the typical uncertainties are 80 G, 
and for HD~56495~B the typical uncertainties are 100 G.  
\citet{Borra1980-magneticAB-survey} surveyed a large number of magnetic Ap stars, 
finding longitudinal magnetic fields of several hundred to several thousand gauss, 
with typical longitudinal fields in the range 500 to 1000 G.  
If such a magnetic field were present in  one of the stars in this study, 
we likely would have detected it well above a $3\sigma$ level. 
\citet{Auriere2007} studied the Ap stars with the weakest magnetic fields, 
finding the weakest stars in their sample reached 100 G longitudinal fields, 
but 300 G longitudinal fields were much more typical even among very weak Ap stars.  
Thus, unless these stars have remarkably weak longitudinal magnetic fields 
(among the $\sim$15 weakest Ap stars known), we would have detected an Ap-type magnetic 
field at $3\sigma$ or better.

\section{Chemical Abundances}
\label{Chemical Abundances}

We performed a detailed abundance analysis of both components of HD~22128 and of HD~56495.  
We did this by directly fitting the binary spectra of the two systems.  
Synthetic spectra were produced with the {\sc Zeeman} spectrum synthesis program 
\citep{Landstreet1988-Zeeman1,Wade2001-zeeman2_etc}, which solves the polarised 
radiative transfer equations assuming Local Thermodynamic Equilibrium (LTE).  
Optimisations to the code for stars with negligible magnetic fields were used 
\citep{Folsom2012-HAeBe-abundances}, since we do not detect magnetic fields 
in these stars.  The observed spectra were iteratively fit using a 
Levenberg-Marquardt $\chi^2$ minimisation routine.  Input atomic data 
for {\sc Zeeman} were extracted from VALD \citep{Kupka1999-VALD}, using an 
`extract stellar' request, with temperatures matching those we find for the stars.  
Model atmospheres were computed with {\sc atlas9} \citep{Kurucz1993-ATLAS9etc}, 
which produces plane-parallel model atmospheres in LTE, and solar abundances 
were used for the calculation of atmospheric structure.  
A grid of model atmospheres was used, spaced at 250 K in \teff\ and at 0.5 
in \lgg, and interpolated to produce the models used in the fitting process.  

Synthetic binary spectra were created by first computing two single star 
spectra with {\sc Zeeman}, in absolute flux units.   Spectra were then Doppler 
shifted by their measured difference in radial velocity, and then added together 
pixel by pixel, weighted by the ratio of stellar radii squared.  
The summed spectrum was then normalised by the sum of 
the continuum spectra of the two stars, weighted by the ratio of radii squared.  
This produces the synthetic binary spectrum.  
Calculating the spectra in absolute units accounts for the $T_{\rm eff}^4$ 
dependence of luminosity, as well as the variation of the two flux distributions 
with wavelength.  \teff\ and $R_{A}/R_{B}$ are both parameters determined 
by the fitting routine.  

The fitting proceeded in the same fashion as described by 
\cite{Folsom2012-HAeBe-abundances} and \citet{Folsom2013-HD98088-Ap-binary}. 
The observed SB2 spectra were fit simultaneously for chemical 
abundances, \vs, microturbulence, \teff, and the ratio of radii.  
If this fit produced well constrained results, as was the case for all 
windows in this study, we attempted to introduce \lgg\ as an additional 
free parameter, to be fit simultaneously with the above parameters.  
If this fit was well constrained, and produced sensible results, 
these results were adopted over previous results that were obtained with a fixed \lgg. 
A fit was considered to be well constrained if it produced values consistent 
with other spectral windows, produced consistent best fit values for different 
initial conditions, and if the synthetic spectrum matched the observation under 
visual inspection.  

Initial values for \teff\ and \lgg\ were estimated from Balmer line profiles 
(H$\alpha$ and H$\beta$), giving initial values for HD~22128~A of 
\teff~$=7500$ K \& \lgg~$=4.0$, for HD~22128~B of \teff~$=6750$ K \& \lgg~$=4.0$, 
for HD~56495~A of \teff~$=7500$ K \& \lgg~$=4.0$, and for HD~56495~B of 
\teff~$=6250$ K \& \lgg~$=4.0$.  These initial estimates were close to the 
final best fit parameters, except for the \teff\ of HD~22128~B, which was 
significantly underestimated.  
Initially a typical  A star microturbulence of 2 \kms\ was used for all stars, 
and solar abundances were initially used for all stars.  
For HD~22128 $R_{A}/R_{B} = 1.2$ was initially used, and for HD~56495 
$R_{A}/R_{B} = 1.4$ was initially used.  

This fitting process for deriving stellar parameters 
was applied first to one star in the binary, 
then to the other star in the binary, then repeated.  
Thus the fitting process for the full SB2 spectrum proceeded in an 
iterative fashion, effectively fitting a single star at each step 
in the process, but ending with a fit to the full SB2 spectrum.  
The results were checked at each step to ensure good quality fits to the 
observation were being achieved and sensible parameters were found.  
The iterative fitting process continued until consistent results were 
found for both stars between subsequent iterations.  
This produced the final best fit parameters for the 
binary system in a given spectral window. 
The same fitting process was used by \citet{Folsom2012-HAeBe-abundances} 
for the SB2 Herbig Be system V380 Ori.  

\begin{figure*}
\centering
\includegraphics[width=4.5in]{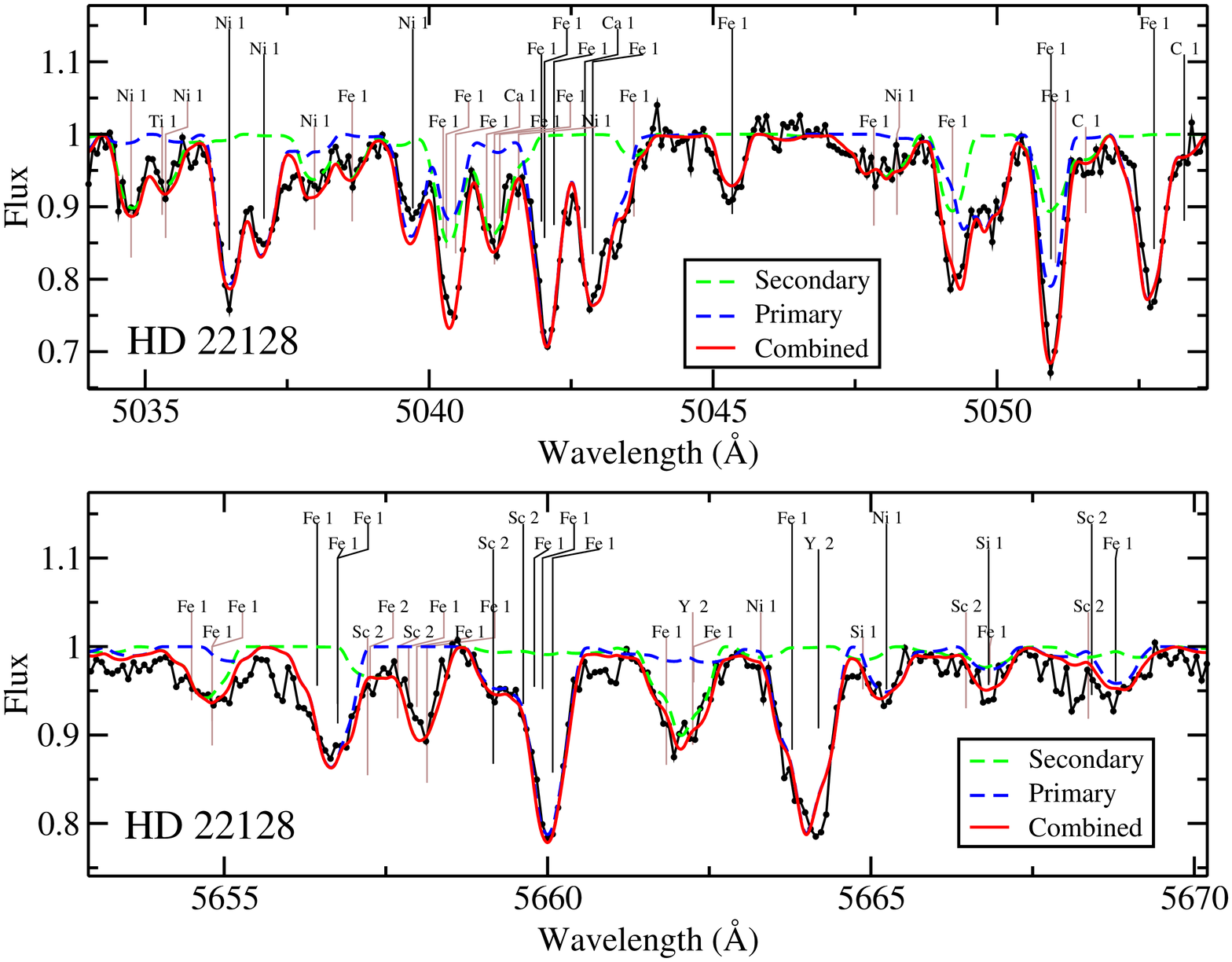}
\includegraphics[width=4.5in]{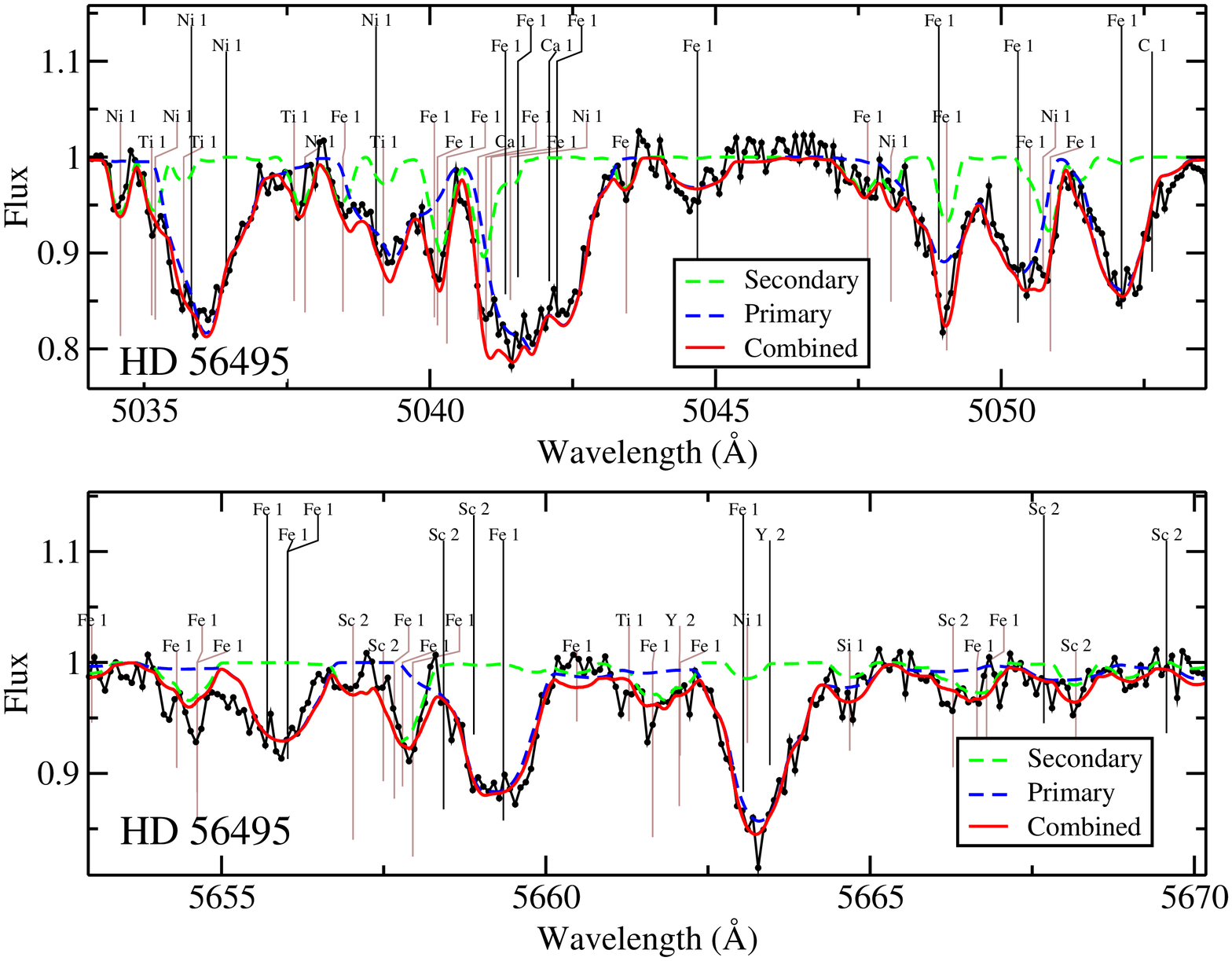}
\caption{Sample fits to the SB2 observations of HD~22128 (top two frames) and 
HD~56495 (bottom two frames).   Fits (smooth lines) to the observations (points) 
show the combined synthetic spectra, as well as the two component synthetic 
spectra.  Major contributing species to the lines are labelled, for both 
components. }
\label{sample-fit}
\end{figure*}

\begin{figure*}
\centering
\includegraphics[width=4.5in]{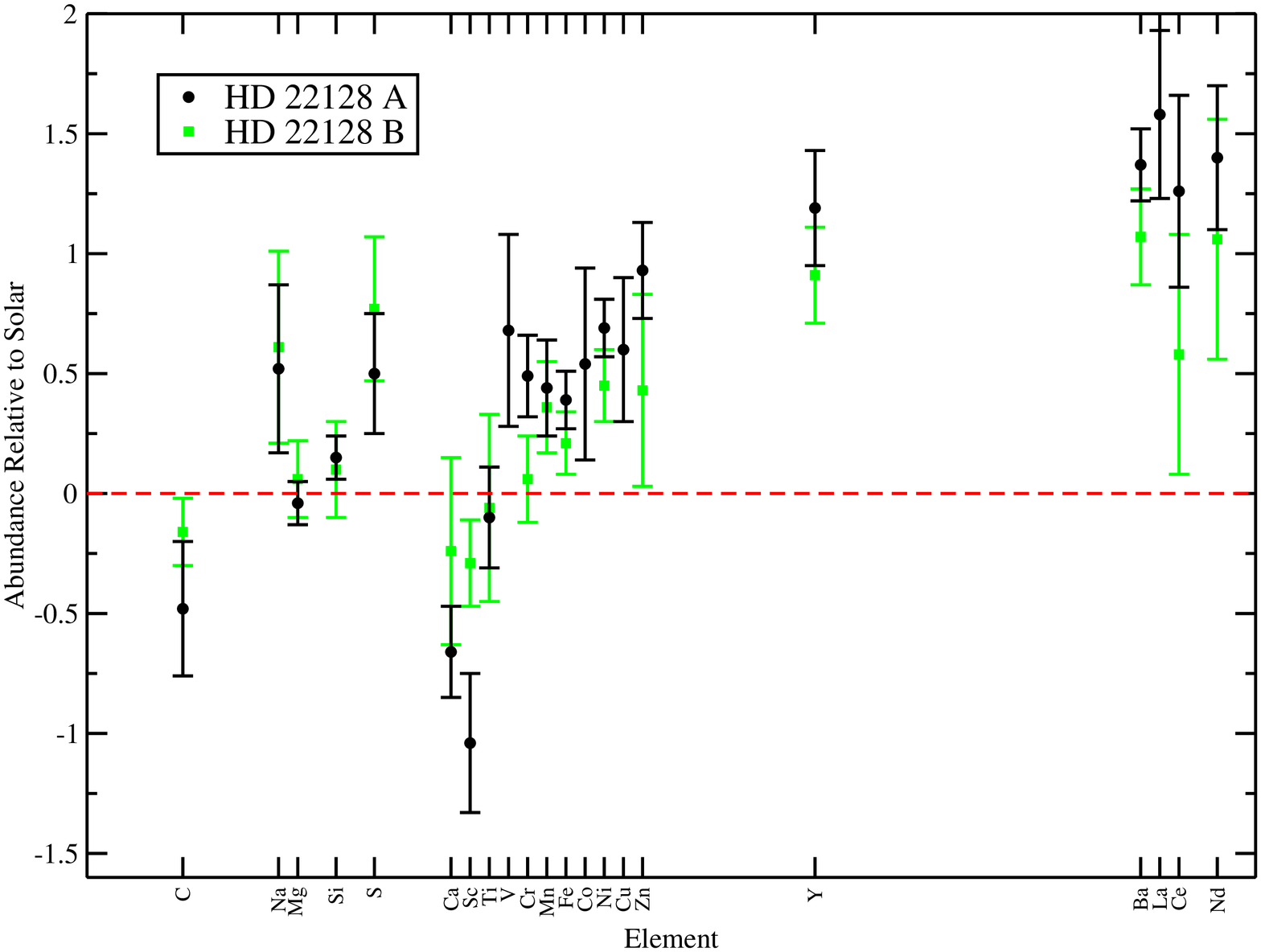}
\includegraphics[width=4.5in]{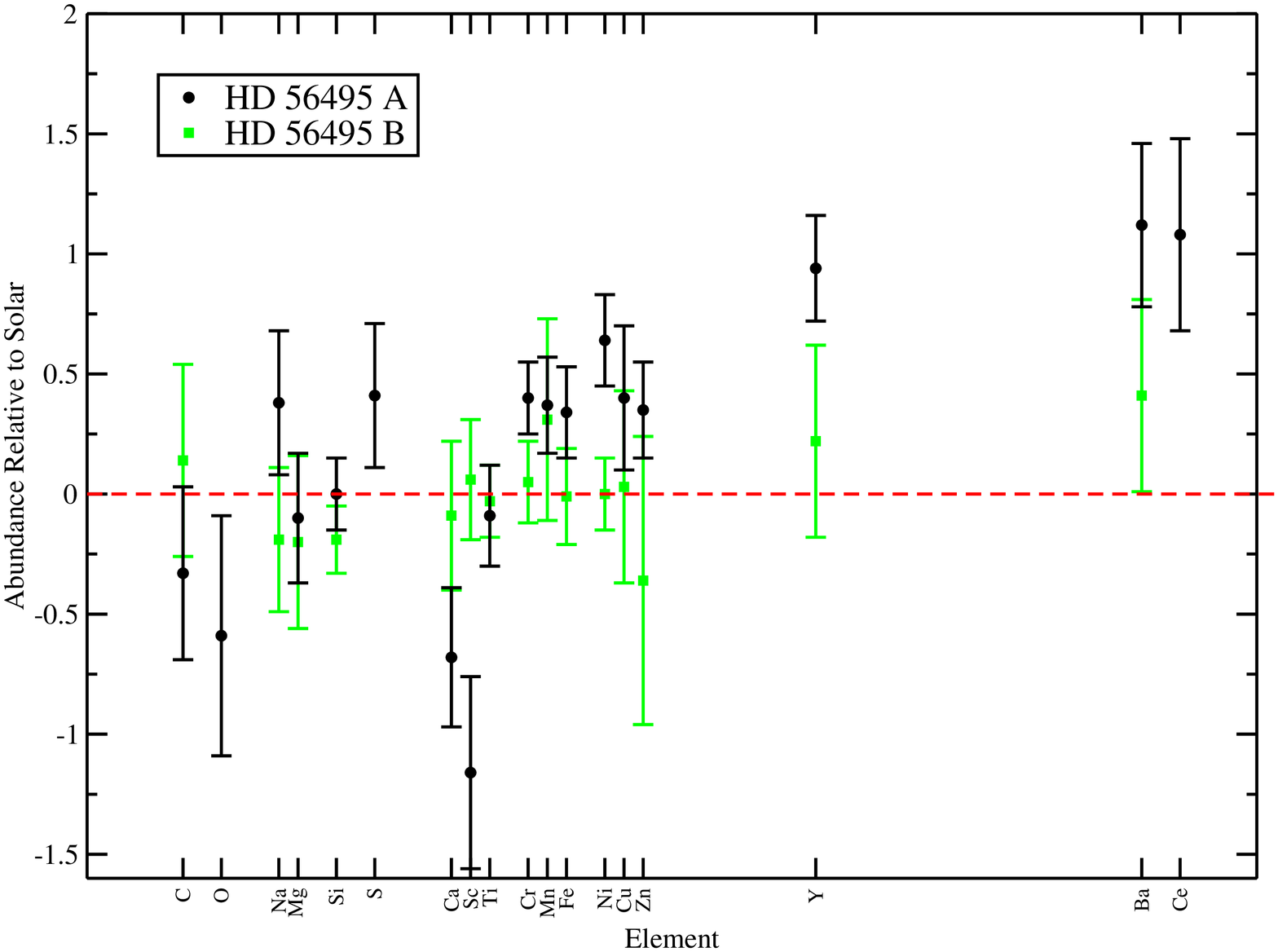}
\caption{Abundances relative to solar for HD~22128~A \& B (top frame) and 
HD~56495~A \& B (bottom frame), averaged over all spectral windows modelled.  
Solar abundances are taken from \citet{Asplund2009-solar-abun}. }
\label{abunplot}
\end{figure*}

In this abundance analysis, the observation of HD~22128 from Feb.~9 2004 
was used, and the observation of HD~56495 from Feb.~1 2004 was used.  
These observations were chosen because the components were widely 
separated in them and they had high S/N.
The fitting process, for all stars, was performed on 4 independent spectral 
windows.  The windows were: 4500-4800 \AA, 5000-5500 \AA, 5500-6000 \AA, and
6000-6520 \AA.  This covers virtually all of the MuSiCoS spectral range, 
excluding Balmer lines.  Sample fits to the observations are presented in 
Fig.~\ref{sample-fit}.  The final best fit values were then taken as the average 
of the best fits from individual windows.  Uncertainties were taken as the 
standard deviation of the best fit values from individual windows.  
For chemical abundances that could only be fit in one or two windows 
an uncertainty was estimated by eye.  This estimate included the scatter 
between lines, noise in the observation, and potential normalisation errors. 
The final best fit values, and their uncertainties, 
are presented in Table \ref{abun-tab}.  The best fit abundances are plotted 
relative to the solar abundances of \citet{Asplund2009-solar-abun} 
in Fig.~\ref{abunplot}.

The fitting process produced a well constrained \lgg\ for only some spectral windows: 
3 windows in HD~56495~A, all 4 in HD~56495~B, and only 1 in HD~22128~A and B. 
For HD~56495 we use the standard deviation of \lgg\ from individual 
windows for the uncertainty in Table \ref{abun-tab}.  For HD~22128, 
the uncertainty on \lgg\ was estimated from the range of values consistent 
with the Balmer lines and the metallic line fit.  

\begin{table*}
\centering
\caption{Best fit chemical abundances and stellar atmosphere parameters for 
both components of HD~22128 and HD~56495.  
Chemical abundance are in units of $\log(N_{X}/N_{tot})$, and abundances fit 
with two or less independent spectral windows are marked with an asterisk (*). 
Solar abundances are from \citet{Asplund2009-solar-abun}.  }
\begin{tabular}{cccccc}
\hline \hline \noalign{\smallskip}
            & HD 22128 A       & HD 22128 B       & HD 56495 A        & HD 56495 B          & Solar\\
\noalign{\smallskip}  \hline \noalign{\smallskip}
\teff\  (K) & $ 7560 \pm  210$ & $ 7480 \pm  310$ & $  7800 \pm  220$ &    $ 6440 \pm  170$ & \\
\lgg        & $ 4.0  \pm  0.5$ & $  4.0 \pm  0.5$ & $  4.0  \pm  0.3$ &    $ 4.2  \pm  0.4$ & \\
\vs\  (\kms)& $ 19.4 \pm  0.9$ & $ 21.0 \pm  1.2$ & $  36.2 \pm  1.8$ &    $ 14.1 \pm  1.3$ & \\
$\xi$ (\kms)& $ 3.59 \pm 0.18$ & $ 3.55 \pm 0.40$ & $  3.44 \pm 0.14$ &    $ 0.93 \pm 0.87$ & \\
$R_{A}/R_{B}$& $ 1.27 \pm 0.13$ &                  & $  1.20 \pm 0.10$ &                     & \\
\noalign{\smallskip}  \hline \noalign{\smallskip}
C      & $-4.05 \pm 0.28$ & $-3.73 \pm 0.14$ & $ -3.90 \pm 0.36$ &    $-3.43 \pm 0.40$*& -3.57 \\
O      & $              $ & $              $ & $ -3.90 \pm 0.50$*&    $              $ & -3.31 \\
Na     & $-5.24 \pm 0.35$ & $-5.15 \pm 0.40$*& $ -5.38 \pm 0.30$*&    $-5.95 \pm 0.30$*& -5.76 \\
Mg     & $-4.44 \pm 0.09$ & $-4.34 \pm 0.16$ & $ -4.50 \pm 0.27$ &    $-4.60 \pm 0.36$ & -4.40 \\
Si     & $-4.34 \pm 0.09$ & $-4.39 \pm 0.20$ & $ -4.49 \pm 0.15$ &    $-4.68 \pm 0.14$ & -4.49 \\
S      & $-4.38 \pm 0.25$ & $-4.11 \pm 0.30$*& $ -4.47 \pm 0.30$*&    $              $ & -4.88 \\
Ca     & $-6.32 \pm 0.19$ & $-5.90 \pm 0.39$ & $ -6.34 \pm 0.29$ &    $-5.75 \pm 0.31$ & -5.66 \\
Sc     & $-9.89 \pm 0.29$ & $-9.14 \pm 0.18$ & $-10.01 \pm 0.40$*&    $-8.79 \pm 0.25$ & -8.85 \\
Ti     & $-7.15 \pm 0.21$ & $-7.11 \pm 0.39$ & $ -7.14 \pm 0.21$ &    $-7.08 \pm 0.15$ & -7.05 \\
V      & $-7.39 \pm 0.40$*& $              $ & $               $ &    $              $ & -8.07 \\
Cr     & $-5.87 \pm 0.17$ & $-6.30 \pm 0.18$ & $ -5.96 \pm 0.15$ &    $-6.31 \pm 0.17$ & -6.36 \\
Mn     & $-6.13 \pm 0.20$*& $-6.21 \pm 0.19$ & $ -6.20 \pm 0.20$*&    $-6.26 \pm 0.42$ & -6.57 \\
Fe     & $-4.11 \pm 0.12$ & $-4.29 \pm 0.13$ & $ -4.16 \pm 0.19$ &    $-4.51 \pm 0.20$ & -4.50 \\
Co     & $-6.47 \pm 0.40$*& $              $ & $               $ &    $              $ & -7.01 \\
Ni     & $-5.10 \pm 0.12$ & $-5.33 \pm 0.15$ & $ -5.14 \pm 0.19$ &    $-5.78 \pm 0.15$ & -5.78 \\
Cu     & $-7.21 \pm 0.30$*& $ \leq -7.4    $*& $ -7.41 \pm 0.30$*&    $-7.78 \pm 0.40$*& -7.81 \\
Zn     & $-6.51 \pm 0.20$*& $-7.01 \pm 0.40$*& $ -7.09 \pm 0.20$*&    $-7.80 \pm 0.60$*& -7.44 \\
Y      & $-8.60 \pm 0.24$ & $-8.88 \pm 0.20$ & $ -8.85 \pm 0.22$ &    $-9.57 \pm 0.40$*& -9.79 \\
Ba     & $-8.45 \pm 0.15$ & $-8.75 \pm 0.20$ & $ -8.70 \pm 0.34$ &    $-9.42 \pm 0.40$ & -9.82 \\
La     & $-9.32 \pm 0.35$*& $              $ & $               $ &    $              $ & -10.90\\
Ce     & $-9.16 \pm 0.40$*& $-9.84 \pm 0.50$*& $ -9.34 \pm 0.40$*&    $              $ & -10.42\\
Nd     & $-9.18 \pm 0.30$*& $-9.52 \pm 0.50$*& $ \leq -9.5     $*&    $              $ & -10.58\\
\noalign{\smallskip} \hline \hline
\end{tabular}
\label{abun-tab}
\end{table*}

\subsection{Chemical abundances of HD 22128 A \& B}

For HD~22128~A we find a well determined \teff\ of $7560\pm210$ K, 
and a somewhat more uncertain \lgg\ of $4.0\pm0.5$. 
For HD~22128~B we find a very similar \teff\ of $7480\pm310$ K and 
\lgg\ of $4.0\pm0.5$.  Through the $\chi^2$ minimisation process 
we also find microturbulences of  $3.59\pm0.18$ \kms\ and 
$3.55\pm0.40$ \kms\ for HD~22128~A and B, respectively.  
These values are somewhat greater than usually seen in A stars, 
but consistent with the elevated values seen in Am stars \citep[e.g.][]{Landstreet2009-vmic2}.
The ratio of radii we find is $R_{A}/R_{B}=1.27\pm0.13$, which provides 
a good fit to the spectra of both stars.  
Under close inspection, the best fit synthetic spectrum does a 
very good job of reproducing the SB2 observation. 
The best fit results are fully consistent with all our observations 
of the HD~22128 system.  

In HD~22128~A we find clear overabundances of the iron-peak elements, 
of $\sim$0.5 dex.  We find strong overabundances of Y, Ba, La, Ce, Nd, 
of 1 to 1.5 dex.   In contrast, Ca and Sc are strongly underabundant, 
at -0.7 and -1.0 dex respectively.  
This pattern of abundances is characteristic of an Am star.  
Na and S appear to be substantially overabundant, 
and C appears to be underabundant.  While the results for C, Na, and S are 
rather uncertain they are consistent with peculiarities seen in other 
Am stars \citep[e.g.][]{Fossati2007}. 

In HD~22128~B we again find clear overabundances of the iron-peak elements, 
Y, Ba, and Nd.  Ce, Na, and S are also marginally overabundant, but very 
uncertain.  Sc is weakly underabundant at -0.3 dex, and Ca may be 
underabundant, though that is not entirely clear.  
This star appears to also be an Am star, 
however not as strongly so as HD~22128~A.

The abundance pattern in HD~22128~B is very similar to HD~22128~A, although 
slightly less extreme.  There are a couple of significant differences.  
Cr appears to be solar in the secondary but overabundant in the primary.  
Sc is substantially closer to solar in the secondary than in the primary 
(by $\sim$0.7 dex), and Ca is likely closer to solar as well.  
The strong similarity in chemical abundances likely reflects 
the similarities in the atmospheric parameters of the stars.  
The effective temperatures, surface gravities, and microturbulences all 
agree to within 1$\sigma$, while \vs\ agrees at slightly over 1$\sigma$.
Thus, if atomic diffusion is giving rise to the  chemical peculiarities in 
these stars, it is likely proceeding in a very similar fashion in both of them. 

\subsection{Chemical abundances of HD 56495 A \& B}

For HD~56495~A we find \teff~$=7800\pm 220$ K and \lgg~$=4.0\pm0.3$, 
and for HD~56495~B we find \teff~$=6440\pm170$ K and \lgg~$=4.0\pm0.4$,
all of which are well constrained by our metallic line fit.  
We find a somewhat elevated microturbulence for HD~56495~A of $3.44\pm0.14$ \kms, 
which is consistent with that seen in Am stars, while for HD~56495~B we find 
a microturbulence of $0.93\pm0.87$ \kms, which is consistent with that seen 
in F stars \citep[e.g.][]{Landstreet2009-vmic2}. 
We find a ratio of radii $R_{A}/R_{B}=1.20\pm0.10$, which is 
consistent with the spectra of both stars.  The best fit 
synthetic spectrum matches the observation very well under close inspection, 
and these results are consistent for all three of our observations of the system

In HD~56495~A we find clear overabundances of iron-peak elements of 
$\sim$0.5 dex, and overabundances of Y, Ba, and Ce of $\sim$ 1 dex.  
There are marginal overabundances of Na and S, though they are very uncertain.  
Ca and Sc are strongly underabundant at -0.7 and -1.2 dex, respectively.  
This is a clear sign of an Am chemical peculiarity.  

In HD~56495~B we find most elements have abundances within 1$\sigma$ of 
solar, and all abundances are well within 2$\sigma$ of solar.  
This star is chemically normal, though the uncertainties on individual 
elements are rather large.  However, if chemical peculiarities were present in the 
secondary with a similar magnitude to the primary, then they would almost 
certainly be detected.  The \teff\ we find for HD~56495~B  of $6440\pm170$ K 
is substantially cooler than one would expect for an Am or Fm star.  

The chemical abundances HD~56495~A in are almost identical to those found in 
HD~22128~A, and also very similar to those found in HD~22128~B. 
The abundances of all elements are within uncertainty of the abundances in 
HD~22128~A, except for Zn which is $2\sigma$ (0.6) dex more abundant in HD~22128~A.  
The abundances are also all within uncertainty of HD~22128~B, except for Sc 
which is more abundant in HD~22128~B (by $2\sigma$, 0.9 dex), and Cr which 
is marginally less abundant in HD~22128~B (by $1.4\sigma$, 0.3 dex).  
The stellar parameters of HD~56495~A are almost identical to those of 
HD~22128~A \& B.  The \teff, \lgg, and microturbulence are all with in 
uncertainty, though \vs\ is 17 \kms\ larger in HD~56495~A.  
Thus, atomic diffusion is likely acting in a very similar way in 
all three of these stars.

\section{Fundamental Parameters}
\label{params}

Both binary systems were observed by the Hipparcos satellite, providing 
precise parallaxes for the systems.  From the reduction by 
\citet{van_Leeuwen2007-Hipparcos_validation}, HD~22128 has a parallax of 
$6.01\pm0.74$ mas ($166\pm20$ pc) and HD~56495 has a parallax of $8.71\pm0.79$
mas ($115\pm10$ pc).  Str\"omgren photometry from \citet{Olsen1994-Stroemgren-photometry} 
was used for HD~22128, and from \citet{Cameron1966-Stroemgren-photometry} 
for HD~56495 \citep[both obtained from the The General Catalogue of 
Photometric Data of][]{Mermilliod1997-GCPD}.  
With this and the bolometric correction from \citet{Balona1994-bolometric-corr} 
we can determine the luminosities of the stars.  
This requires the ratio of luminosities of the stars in the systems, 
which we can calculate from the ratio of $T_{\rm eff}^4$ 
and ratio of radii squared.  The bolometric corrections for the 
components in each system are very similar.  
Since the stars are nearby, interstellar extinction was neglected. 
The uncertainty in luminosity is dominated by uncertainties in 
the ratio of luminosities and the distance, though uncertainties 
in the bolometric correction and photometry are included.  
The luminosities derived are presented in Table \ref{fundimental-param-table}.

\begin{figure}
\centering
\includegraphics[width=3.3in]{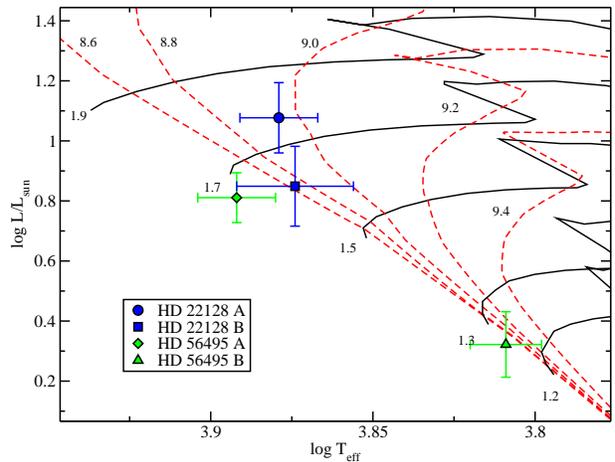}
\caption{H-R diagram for HD 22128 and HD 56495.  Evolutionary tracks (solid lines) 
and isochrones (dashed lines) are shown, labelled 
by mass in solar masses and $\log {\rm age}$, respectively. 
Evolutionary tracks are from \citet{Schaller1992-ms-evol} for standard 
mass loss and Z=0.02.  }
\label{hr-diagram}
\end{figure}

\begin{table}

\centering
\caption{Derived fundamental parameters for the HD~22128 and HD~56495 systems. }
\scriptsize
%\begin{tabular}{l@{ }c@{ \ }cc@{ \ }c}
\begin{tabular}{lc@{ \ }cc@{ \ }c}
\hline\hline
               & HD 22128 A    & HD 22128 B    & HD 56495 A    & HD 56495 B    \\
%               & \multicolumn{2}{c}{HD 22128} & \multicolumn{2}{c}{HD 56495}   \\
%               &      A        &       B       &      A        &       B       \\
\hline                                                 
$m_{v}$   &\multicolumn{2}{c}{$7.603\pm0.005$}&\multicolumn{2}{c}{$7.680\pm0.013$}\\
$d$ (pc)       &\multicolumn{2}{c}{$166\pm20$} &\multicolumn{2}{c}{$115\pm10$} \\
\teff\ (K)     & $7560\pm210$  & $7480\pm310$  & $7800\pm220$  & $6440\pm170$  \\
$L\ (L_{\odot})$& $11.9\pm3.2$  & $7.1\pm2.2$   & $6.5\pm1.2$   & $2.1\pm0.5$ \\
$R\ (R_{\odot})$& $2.01\pm0.29$ & $1.58\pm0.27$ & $1.39\pm0.15$ & $1.16\pm0.16$ \\
$M\ (M_{\odot})$& $1.77\pm0.09$ & $1.61\pm0.09$ & $1.67\pm0.07$ & $1.25\pm0.06$ \\
$\log$ age     &\multicolumn{2}{c}{$8.95^{+0.08}_{-0.15}$}&\multicolumn{2}{c}{$<8.6$}\\
$i_{orb}\ (\degr)$&\multicolumn{2}{c}{$49.7\pm1.2$}&\multicolumn{2}{c}{$83.8\pm5.2$}\\
$a_{orb}\ (R_{\odot})$&$9.0\pm0.2$& $9.7\pm0.2$  &$23.8\pm0.5$   & $31.0\pm0.6$  \\
\lgg\ (cgs)    & $4.08\pm0.13$ & $4.25\pm0.15$ & $4.37\pm0.10$ & $4.40\pm0.12$ \\
\hline\hline  
\end{tabular} 
\label{fundimental-param-table} 
\end{table}

With \teff\ and luminosity known, the stars can be placed on the Hertzsprung Russell 
(H-R) diagram, as shown in Fig. \ref{hr-diagram}.  By comparing with evolutionary 
tracks and isochrones from \citet{Schaller1992-ms-evol}, calculated with 
standard mass loss and Z=0.02, we can derive masses and ages for the stars.  
HD~56495~B sits on the zero age main sequence (ZAMS) line, and HD~56495~A 
sits nearly $1\sigma$ below the ZAMS, consequently we can only place upper limits 
on the ages of these stars.  It is possible that unusually large extinction 
is responsible for the low position of HD~56495~A on the H-R diagram, 
but it could simply be due a $1\sigma$ error in a stellar parameter.  
HD~22128~B also appears to sit on the ZAMS, so we assume it has the same ages as
HD~22128~A.  Ages and masses are presented in Table \ref{fundimental-param-table}.

We can use the derived masses, together with the dynamical $M \sin^3 i$ from 
\citet{Carrier2002-binarity_in_Ap} to derive a value for the orbital inclination 
$i_{orb}$.  The value of $i_{orb}$ for HD~22128 is well constrained, 
the value for HD~56495 ($i_{orb} = 83.8\pm5.2$) is also fairly precise, 
but near $90\degr$.  
We can then use the orbital inclination to derive the semi-major axes 
of each star's orbit $a_{orb}$, based on the dynamical $a \sin i$ from \citet{Carrier2002-binarity_in_Ap}.
We can also use the derived masses from the H-R diagram and stellar radii from the 
luminosities and \teff\ to make an alternate measurement of \lgg.  
These results are presented in Table \ref{fundimental-param-table} and are 
fully consistent with the spectroscopic \lgg\ values, but have significantly 
smaller formal uncertainties. 

The mass ratios of the stars from the H-R diagram agree well with the 
dynamical mass ratios of \citet{Carrier2002-binarity_in_Ap}.  
For HD~22128 these are $M_{A}/M_{B} = 1.10\pm0.08$ (H-R diagram) and 
$M_{A}/M_{B} = 1.08\pm0.02$ (dynamical), and for HD~56495 they are 
$M_{A}/M_{B} = 1.34\pm0.09$ (H-R diagram) and $M_{A}/M_{B} = 1.30\pm0.06$ (dynamical). 
This provides some confidence in the H-R diagram positions of the stars. 

Whether the stars are tidally locked was investigated, since they are in 
very close binary systems.  If we assume tidal locking, then the rotational 
period is equal to the orbital period.  This period combined with the stellar 
radius can be used to derive a hypothetical equatorial rotational velocity.  
For HD~22128~A, this hypothetical value is $20.0\pm2.9$ \kms, 
and for HD~22128~B it is $15.7\pm2.7$ \kms.  
For HD~22128~A the observed \vs\ ($19.4\pm0.9$ \kms) is consistent 
with this hypothetical equatorial rotational velocity, 
thus the star could be tidally locked as long as the rotational axis has an 
inclination of $76 ^{+14}_{-22}\degr$.  However, this would require the 
rotational and orbital axes to be misaligned by a little over $1\sigma$. 
For HD~22128~B, the observed \vs\ ($21.0\pm1.2$ \kms) is inconsistent with 
the hypothetical equatorial value, thus the star is likely not tidally locked. 
The stellar radius of HD~22128~B would have to be 1.4 times larger 
for \vs\ to be consistent with tidal locking, which at $2.3\sigma$ 
is unlikely but not impossible.  
Alternately, if we assume the orbital and rotational axes are aligned,  
we can use \vs\ to estimate the rotational periods.  For HD~22128~A this 
hypothetical rotational period is $4.0\pm0.6$ days, and for HD~22128~B 
it is and $2.9\pm0.5$ days.  Thus HD~22128 has probably experienced a 
large amount of tidal braking, and the system is close to being tidally 
locked, but does not appear to be actually tidally locked.  

We can perform the same evaluation of tidal locking for HD~56495.  
Assuming the rotational and orbital periods are equal, the hypothetical 
equatorial rotational velocities are $2.6\pm0.3$ \kms\ for HD~56495~A, 
and $2.2\pm0.3$ \kms\ for HD~56495~B.  These are completely inconsistent 
with the observed \vs\ for both HD~56495~A ($36.2\pm1.8$ \kms) and 
HD~56495~B ($14.1\pm1.3$ \kms).  Alternately, assuming the rotational 
and orbital axes are parallel, we find the hypothetical rotational 
periods of $1.93\pm0.24$ days for HD~56495~A, and $4.15\pm0.68$ days 
for HD~56495~B.  The \vs\ values are relatively low, 
and thus the HD~56495 system may have experienced significant tidal braking, 
but the stars are clearly not tidally locked.

\section{Discussion and Conclusions}

\citet{Carrier2002-binarity_in_Ap} derived some tentative stellar parameters 
for HD~22128 and HD~56495, using the assumption that the two components in each 
system were identical.  They used photometry to estimate \teff, and parallaxes 
to estimate absolute luminosities.  For HD~22128 they found 
\teff~$=6900$ K, \lgg~$=3.65$, $\log L/L_{\odot} = 0.95$, and 
$R/R_{\odot} = 2.10$, leading to $M = 1.99$ $M_{\odot}$ and 
$i_{orb} \sim 48 \degr$.  These results are consistent with our values at 
between 1 and 2$\sigma$, which is acceptable given the assumptions involved.  
For HD~56495 they found \teff~$=7179$ K, \lgg~$=4.00$, $\log L/L_{\odot} = 0.77$, 
and $R/R_{\odot} = 1.58$, leading to $M = 1.80$ $M_{\odot}$ and $i_{orb} \sim 75 \degr$.  
These values are inconsistent with ours, since the components of HD~56495 
have significantly different temperatures and masses, however they do fall 
between our values, except for radius and mass.  
A few other large surveys have also derived fundamental parameters for 
HD~22128 or HD~56495 \citep[e.g.][]{Masana2006-stellar-param-from-2MASS,Holmberg2009-Geneva-Copenhagen3,Bailer-Jones2011-stellar-param-estimation,Casagrande2011-Geneva-Copenhagen-rered}, 
however they neglected the binary nature of the systems and thus derived 
inaccurate parameters.  \citet{Nordstrom2004A&A-Geneva-Copenhagen1} considered  
HD~22128 as a spectroscopic binary and found a mass ratio of $0.927\pm0.005$, 
which is inconsistent with the result of \citet{Carrier2002-binarity_in_Ap}, 
although this appears to be based on very few radial velocity measurements.  

\citet{Decin2003-vega-type-survey} identified HD~22128 as a main sequence 
star with a debris disk, or a Vega analogue, based on an infrared excess.   
However, they did not recognize the star as a binary, 
or an Am star, which brings the result into question.  
\citet{Wyatt2007-Debris-Disks-survey} also come to the same conclusion that 
the star has a debris disk, again not considering binarity and peculiarity, 
however they note that the system is anomalous and their luminosity or age 
may have been miscalculated.  Since both components of HD~22128 have similar 
temperatures one would not expect a large infrared excess when treated 
as a single star, so it is possible that there is a debris disk 
around the system.  IR photometry from the Wide-field Infrared Survey Explorer 
(WISE) satellite \citep{Wright2010-WISE-general}
shows a far IR excess for HD~22128, as compared to HD~56495, 
in the W3 (11.6 $\mu$m) and W4 (22.1 $\mu$m) bands.   
The stars agree well in the shorter wavelength W1 and W2 bands, and in 
the Two Micron All Sky Survey \citep[2MASS;][]{Cutri2003-2MASS-point-sources} 
IJK photometry, which lends some support to the presence of a debris disk.

\citet{Koen2002-hipparcos-var} report periodic variability in the 
Hipparcos photometry of HD~22128.  The variability has a period of 0.10545 days 
and an amplitude of 0.0092 magnitudes.  This period is inconsistent with 
the orbital or any plausible rotational period.  However, the period 
and amplitude are roughly consistent with $\delta$ Scuti pulsations.  
Interestingly, low amplitude $\delta$ Sct and $\gamma$ Dor pulsations appear 
to occur frequently among Am stars 
\citep{Smalley2011-Am-puslations,Balona2011-Am-puslations}.
The interaction between Am chemical peculiarities and pulsations could 
provide valuable insights into both phenomena.  

Since the stars are in close binary systems, 
and HD~56495 has an orbital inclination approaching $90\degr$, 
it is worth checking whether the systems could be eclipsing.  
Using the semi-major axes, stellar radii, and inclinations we find HD~22128 
could not be eclipsing.  HD~22128 would have to have an orbital 
inclination of $\geq 79\degr$ to show a partial eclipse.  
HD~56495 would not be eclipsing at our best inclination of 
$i_{orb}=83.8\degr$ but there is enough uncertainty that $i_{orb}$ 
could be $90\degr$.  Examining the Hipparcos photometry of both systems, 
phased with their orbital periods, shows no coherent variability.  
Thus there is no evidence that either system is eclipsing.  
This observation places an upper limit on the orbital inclination 
of HD~56495 of $i_{orb} < 87.8\degr$.

Our magnetic and spectroscopic measurements of HD~22128 strongly indicate that 
neither component of the system is an Ap star.  If a magnetic field typical 
of an Ap star (maximum $B_{l}$ typically 500 to 1000 G; 
\citealt{Borra1980-magneticAB-survey}) were present in either 
star, we almost certainly would have found it.  However, the primary of 
the system is a clear Am star, and the secondary appears to be a somewhat 
weaker Am star.  The symmetry of the lines profiles of both components, 
and the lack of variability in the spectra other than orbital motions, 
both support the conclusion that neither star is an Ap star.  
The stars in the system are close to being tidally locked, 
but probably are not quite synchronised yet.  The classification of 
HD~22128 as an Ap star rests on the work of 
\citet{Abt1979-classification-cp}, who only had classification 
resolution spectra.  It seems likely that they misclassified the system as 
an Ap star rather than two Am stars as a consequence of unrecognized 
binarity, due to low spectral resolution and perhaps an unfortunate orbital phase.  

HD~56495 is clearly not an Ap star, based on our spectroscopic and 
magnetic measurements.  If a typical Ap star magnetic field were present 
in either component it would most likely have been detected.  
We find the primary of HD~56495 is a clear Am star, and the secondary 
appears to be a chemically normal F star.  
The classification of the system as an Ap star rests on the magnetic 
measurement and spectral classification of \citet{Babcock1958-magnetic-catalog}.  
If the magnetic field of $B_{l} = +570 \pm 200$ G reported by 
\citet{Babcock1958-magnetic-catalog} was present, it is very likely we would 
have detected it.  It is much more likely that \citet{Babcock1958-magnetic-catalog}
slightly underestimated the uncertainties, or simply over-interpreted 
a measurement that is only significant at $2.8\sigma$.   
Thus any claims of a magnetic detection are likely spurious.  
Conversely, the spectroscopic identification of the system as an Am star by
\citet{Bertaud1967-Am-classifications} is correct, at least for the primary. 

Our magnetic non-detections are consistent with larger surveys of magnetic 
fields in Am stars \citep[e.g.][]{Shorlin2002}.  A recent study by 
\citet{Auriere2010-survey-magnetic-Am-HgMn} on a sample of 12 Am stars placed very strong 
limits on the presence of magnetic fields with 1$\sigma$ uncertainties of 1 to 3 G.  
However, \citet{Petit2011-siriusA-mag} detected the presence of a magnetic field 
in the Am star Sirius A, with an extremely small longitudinal strength of $0.2 \pm 0.1$ G.  
This appears to be similar, at least in strength, to the sub-gauss field detected 
in Vega by \citet{Lignieres2009-Vega-magnetic}.  
While the magnetic field in Sirius A is likely not dynamically significant, 
and probably is not important for atomic diffusion, its presence is striking. 
This poses the question of whether such extremely weak fields are common among Am stars, 
or A stars in general, or are these unique cases?
The presence of such weak fields in HD~22128 and HD~56495 cannot be ruled out 
by our observations.  Regardless, these extremely weak magnetic fields 
are clearly distinct from those seen in Ap stars. 

HD~22128 and HD~56495 represent interesting cases of the development of Am peculiarities 
from a theoretical standpoint.  In HD~22128, the stars are in a very close binary, 
and thus tidal interactions can easily slow the rotation rates of the star, 
nearly to the point of synchronisation with the orbital period.  This reduces 
meridional circulation, allowing helium settling to occur and atomic diffusion 
to proceed efficiently, producing Am peculiarities.  The two stars have very 
similar parameters, thus atomic diffusion likely proceeds in a very similar 
fashion in both stars, producing very similar observed peculiarities.  
HD~56495 is also a close binary, and thus the stars have likely undergone 
significant tidal braking, though perhaps not to the same extent as HD~22128 
since the system has a wider separation and may be younger.  The primary of 
HD~56495~A has a very similar \teff\ and \lgg\ to the stars in HD~22128, and 
displays very similar chemical peculiarities.  The secondary of HD~56495 is 
substantially cooler, falling outside of the typical temperature range of 
Am or Fm stars, and likely has large enough convection zones to inhibit 
atomic diffusion.  Thus the chemical abundances seen in these stars can be well 
explained by the combination of tidal braking and atomic diffusion. 

The status of these stars as Am stars, not Ap stars, has an important 
impact on the incidence of Ap stars in close binaries.  
Evidently Ap stars are even more rare in close binary systems than previously 
realised.  This leaves only three known or proposed Ap stars in an SB2 system with 
an A star (HD~55719, HD~98088, and HD~5550), and two more Ap stars in SB2 
systems with G giants (HD~59435, HD~135728).  
The status of HD~98088 as containing an Ap star was well established by 
\citet{Folsom2013-HD98088-Ap-binary}, and the two systems containing 
a giant and an Ap were also well established by \citet{Wade1999-hd59435-Ap-binary} 
and \citet{Freyhammer2008-new-Ap+1binary}.  HD~55719 was established by 
\citet{Bonsack1976-hd55719-Ap-binary} who made repeated magnetic measurements, 
however modern observations are warranted.  HD~5550 has no magnetic measurements, 
nor any modern spectroscopic study, and thus is need of new observations 
to confirm or refute its status as an SB2 system with an Ap star.

The BinaMIcS collaboration plans a couple follow-up observations of 
HD~22128 and HD~56495 with the ESPaDOnS spectropolarimeter at 
the Canada France Hawaii Telescope.  These observations should place 
significantly smaller limits on the magnetic fields of these stars.  
However, in light of the new results in this paper, they are not likely 
to be major targets for the BinaMIcS project.  
Observations of the other SB2 systems possibly containing Ap stars, 
as discussed here, are also planned.  

\section*{Acknowledgements} 
GAW is supported by an Natural Science and Engineering Research Council (NSERC Canada) Discovery Grant.

\bibliography{massivebib.bib}{}
\bibliographystyle{mn2e}
%This is an unofficial fix of mn2e.bst for articles with long co-author lists
%\bibliographystyle{mn2e-williams}

\label{lastpage}

\end{document}